\documentclass{optica-article}

\journal{opticajournal} % for journals or Optica Open

\articletype{Research Article}

\usepackage{lineno}
\usepackage{bm}
\begin{document}

\title{Structured detection beyond spatial imaging}

\author{Bruno K. Nowak\authormark{*} and Shu-Wei Huang}

\address{Department of Electrical, Computer, and Energy Engineering, University of Colorado, Boulder, Colorado 80309, USA}

\email{\authormark{*}bruno.nowak@colorado.edu} %% email address is required; see note below about the corresponding author designation

% use {asbstract*} to suppress the copyright line. Copyright information will be added in production

%\begin{abstract*} 
%\end{abstract*}

%%%%%%%%%%%%%%%%%%%%%%%%%%  body  %%%%%%%%%%%%%%%%%%%%%%%%%%
\begin{abstract*} 

In a point-scanning microscope, detection through a pinhole improves resolution and replacing the pinhole with a detector array provides redundant encoding of the object spectrum, enabling correction of nonidealities in the imaging system or further resolution improvements, for instance. Here, we show that the mathematical principles underlying structured detection are not intrinsically spatial and therefore structured detection may be used for imaging distributions in other physical domains, such as time, frequency, and velocity space. We numerically illustrate this generality by applying structured detection to an electromagnetic distribution in time using four-wave mixing (FWM). The resulting image reconstructions outperform the temporal equivalent of confocal detection, even as the FWM models are made more realistic and shot noise is considered. 

\end{abstract*}

\section{Introduction}

Confocal microscopy provides an early example of the benefits of sampling an object with an additional localized response: introducing the finite detection point-spread function associated with the pinhole expands frequency support and consequently improves resolution \cite{McCutchen1967}. Soon after demonstrating the confocal microscope, researchers replaced the pinhole with a detector array to sample the object from different detector positions for each excitation beam position, and these redundant measurements were shown to improve recovery of some spatial frequencies compared to confocal detection \cite{BerteroPike1982,Bertero1987,Walker1993}. Since then, the redundancy has been leveraged to improve spatial resolution \cite{Sheppard1988,Muller2010,Lu2009,Lu2013,Kuang2016}, characterize and correct the imaging system \cite{Kuang2016,Fersini2025,Lim2026}, and reduce the spatial sampling density required for image reconstruction \cite{Zunino2023}.

The frequency-support expansion and redundant encoding are not intrinsically spatial concepts, yet they have been exploited almost entirely in the context of conventional spatial imaging. In this paper, we show these features follow from a general class of measurements in which a distribution is sampled by multiple translated kernels. Because the derivation depends neither on what the distribution represents nor the physical domain over which it is defined, the frequency-support expansion and redundant encoding inherent to the measurement structure extend to other physical domains. To further illustrate this extension, we numerically show structured detection of an electromagnetic distribution (EMD) in time via FWM and compare images produced by this modality with those generated by the temporal equivalent of confocal detection. 

\section{Theory}
The data acquired from a descanned point-scanning microscope with the detector element and object confined to the same single dimension can be modeled as
\begin{equation}
I(x_0,x_1)\propto\int \rho_0(x)h_{\mathrm{exc}}(x-x_0)h_{\mathrm{det}}(x-x_1)dx,
\label{eq:1}
\end{equation}
where $x$ denotes the object-space coordinate, $x_0$ the coordinate of the center of the excitation beam on the object, $x_1$ the coordinate of the center of the detector element imaged back to object space, and $\rho_0$ the object; $h_{\mathrm{exc}}$ and $h_{\mathrm{det}}$ are integral-normalized excitation and detection point spread functions (PSFs) corresponding to 1D slices through their respective 3D PSFs. 

More generally, equation~(\ref{eq:1}) is a member of the family of linear measurement operators
\begin{equation}
(\mathcal{M}_N\rho)(\lambda_0,\ldots,\lambda_{N-1}) \equiv \int \rho(\zeta) \prod_{i=0}^{N-1}h_i(\zeta-\lambda_i)d\zeta.
\label{eq:2}
\end{equation}
Here, $\zeta$ denotes the coordinate over which the distribution $\rho$ is defined, and $\lambda_i$ specifies the center position of the $i$th kernel $h_i$. In the Fourier domain, the N-kernel measurement operator for structured detection (N-MOSD) defined by equation~(\ref{eq:2}) can be written as
\begin{equation}
\begin{aligned}
(\tilde{\mathcal{M}}_N\tilde{\rho}) (k_0,\ldots,k_{N-1}) 
= \tilde{\rho} \left( \sum_{i=0}^{N-1}k_i \right) \prod_{i=0}^{N-1} \tilde{h}_{i}(-k_i).
\label{eq:3}
\end{aligned}
\end{equation}
This result shows that a distribution frequency $q=\sum_i k_i$ is accessible whenever the transfer function $\tilde{h}_i$ of every kernel is nonzero at the corresponding $k_i$, and therefore the accessible distribution-frequency range is the Minkowski sum of the individual kernel supports. Furthermore, an N-MOSD with more than one kernel and independently tunable kernel centers provides redundant encoding of the distribution spectrum. 

Importantly, no assumptions were required about what the distribution represents or the physical domain over which it is defined to obtain these results. Therefore, the frequency-support expansion enabled by sampling a distribution with additional kernels is not intrinsically restricted to spatial imaging. Similarly, the applications of redundant encoding that have been developed for imaging EMDs in the spatial domain are, in principle, applicable to distributions in other domains as well. Indeed, the generalized N-MOSD with independently tunable kernel centers can help separate systematic noise from measurement data; extract kernels to, for instance, deconvolve the distribution; and potentially improve resolution of the distribution without introducing deconvolution protocols, as happens in image scanning microscopy or virtually structured detection (see section 1 of the SI).

\section{Simulation}

To illustrate structured detection outside of the spatial domain, we numerically study pulsed FWM followed by temporally integrated detection of the idler. The redundant encoding enabled by the independently tunable delays of the two input gates is used to remove nuisance and deconvolve the distribution from the extracted gate profiles for more accurate reconstruction of the distribution compared to confocal detection, where the two gates are perfectly overlapped while they sample the distribution.  
 
More specifically, we study the integrated idler intensity for increasingly realistic models of FWM (see section 2A of the SI). The first model assumes negligible depletion of the input fields and excludes propagation effects besides nonlinear conversion of the idler, producing an integrated idler intensity that is the form of a 2-MOSD. The second model breaks this ideal form by adding group-velocity mismatch (GVM) to the first model, and the third model further adds group-velocity dispersion (GVD), phase mismatch, self- and cross-phase modulation, and nonnegligible depletion of the input fields to the second.

For enacting the FWM process, we choose a $1$-cm-long, $1.45$-$\mathrm{\mu m}$ wide, and $0.725$-$\mathrm{\mu m}$ tall $\mathrm{Si_3N_4}$ waveguide with $\mathrm{SiO_2}$ cladding (see section 2B of the SI). The aforementioned cross-section dimensions provide a minimal phase mismatch of $\Delta k =-7.10\ \mathrm{m^{-1}}$ between the fundamental TE-like modes at $1550$, $775$, $800$, and $1458.82$ nm, which are the center wavelengths of the first gate, second gate, distribution, and idler, respectively. At those corresponding wavelengths, the GVD is $-86.54$, $117.08$, $103.94$, and $-93.03$ $\mathrm{fs^2/mm}$, and the GVM relative to the idler is $-78.80$, $-69.21$, $5.20$, and $0$ $\mathrm{fs/cm}$.

The output idler energy is determined as follows. We assume that the input slowly-varying envelopes of the first and second gates take the form of a $\mathrm{sinc}$ function with a center-to-first-zero width of $T_0=100\ \mathrm{fs}$ and amplitude of $\sqrt{P_0}=\sqrt{20\ \mathrm{W}}$, while the distribution power profile is the sum of two Gaussians separated by $d=0.9T_0$ with a standard deviation of $\sigma=20\ \mathrm{fs}$ and the same amplitude as the gates. These input fields are propagated through the waveguide using the symmetrized split-step Fourier method \cite{Agrawal}, with each propagation step consisting of a linear half-step, a full nonlinear step based on a fourth-order Runge--Kutta method, and a second linear half-step (see section 2C of the SI). At the output of the waveguide, the detected idler energy is contaminated with nuisance $B(t_1)$ corresponding to the cross-correlation between the first gate and distribution at the input such that
\begin{equation}
B(t_1) = \eta \int |A_3(0,\tau)|^2 |A_1(0,\tau;t_1)|^2 \,d\tau,
\end{equation}
where $t_1$ denotes the delay of the first gate, $A_1$ and $A_3$ respectively denote the envelopes of the first gate and distribution, and $\eta$ sets the nuisance amplitude. The nuisance amplitude is exaggerated for demonstration purposes and chosen so that the peak nuisance energy is equal to the peak idler energy:
\begin{equation}
\max_{t_1} B(t_1) = \max_{t_1,t_2} \int |A_4(L,\tau;t_1,t_2)|^2 \,d\tau,
\end{equation}
where $t_2$ is the delay of the second gate and $L$ the length of the waveguide. In one set of measurements, the contaminated idler is detected directly without considering shot noise. In a second set, the idler and nuisance are attenuated so that the peak detected count rate for any given pair of gate delays is $10$ Mcps, and this normalization is forced across all three models. Shot noise is then introduced by drawing the detected photon number at each gate-position pair from a Poisson distribution with mean
\begin{equation}
\mu(t_1,t_2)=R(t_1,t_2)T,
\end{equation}
where $R(t_1,t_2)$ is the attenuated detected count rate and $T$ is the integration time at each gate-position pair, varying between $10$ and $100\ \mathrm{ms}$.

Two sets of gate-position pairs are considered. In one, the first gate samples a $600\ \mathrm{fs}$ window centered on the distribution with a $25\ \mathrm{fs}$ step size, and the second gate samples the same window about each position of the first gate. This configuration closely satisfies the Nyquist sampling requirements of the measurement and is associated with $25\times25=625$ gate-position pairs, which is likely to be viable experimentally. In the second set, the scan window is extended to $1200$ fs and the step size is reduced to $6.25$ fs to produce $193\times193=37249$ gate-position pairs. This scan is not as experimentally feasible, but, nevertheless, it is included to probe reconstruction performance under nearly ideal experimental conditions.

The distribution is reconstructed from the resulting measurement data in an algorithm involving only linear least-squares problems (see section 2D of the SI). Briefly, in each iteration, the nuisance is estimated and then removed from the measurement data, after which the gate profiles are estimated and used to deconvolve the distribution; this process is then repeated until convergence. The only prior knowledge of the system entering the algorithm is the cutoff frequency of the gates and dimensional structure of the nuisance. 

\section{Results}

\begin{figure}[h]
    \centering
    \includegraphics[width=\linewidth]{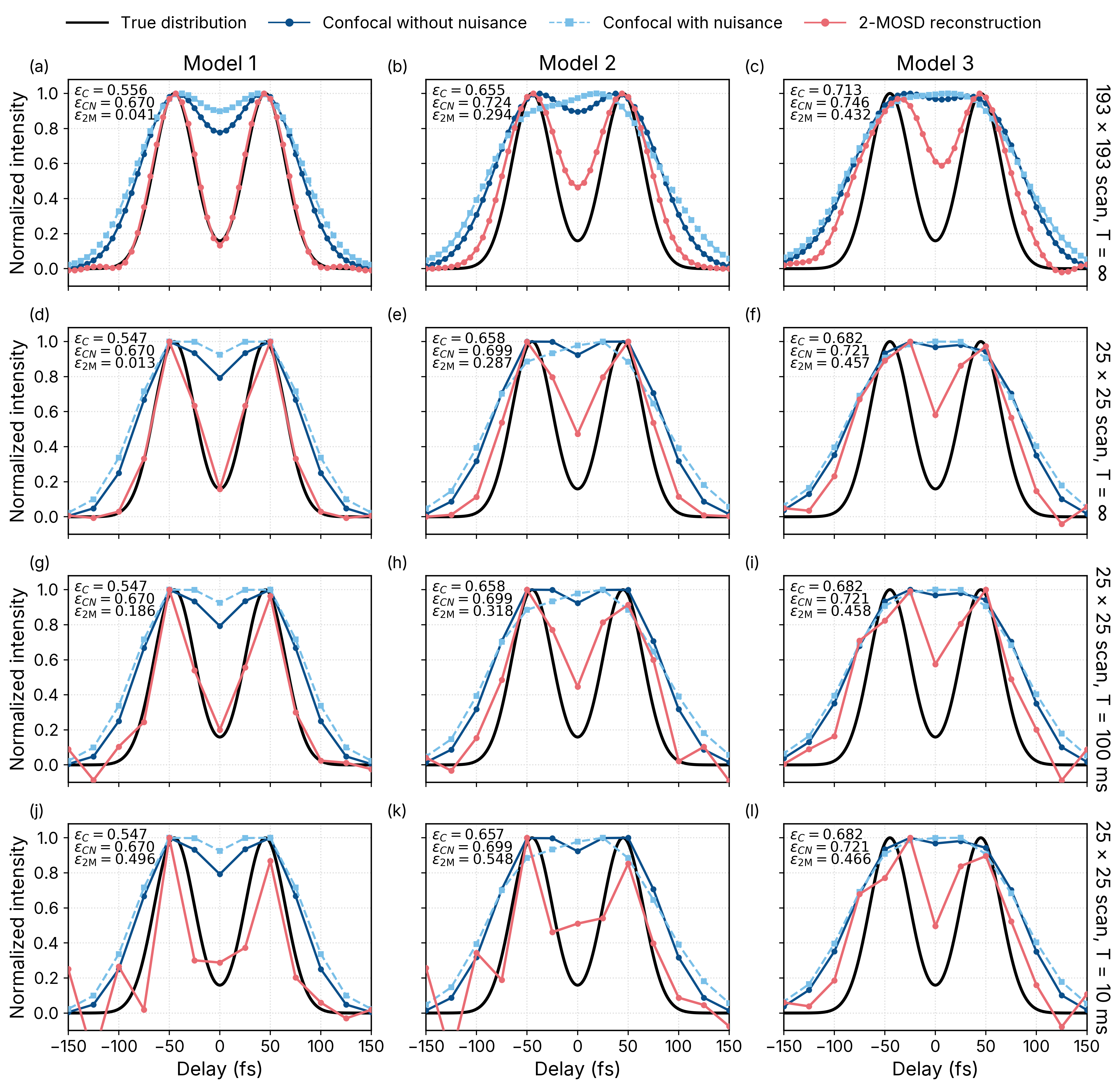}
    \caption{
    Reconstruction of temporal distribution consisting of two Gaussian peaks with $\sigma=20~\mathrm{fs}$ separated by $d=90~\mathrm{fs}$ for the three FWM models (columns) and four different gate-scan configurations (rows). The true distribution (black), 2-MOSD reconstruction (red), and confocal reconstructions with (light blue) and without (dark blue) nuisance are shown after normalization and temporal alignment with the true distribution. The respective relative 2-norm errors $\epsilon_{\mathrm{2M}}$, $\epsilon_{\mathrm{CN}}$, and $\epsilon_{\mathrm{C}}$ are given in the upper-left corner of each panel.
    }
    \label{fig:fwm_6panel}
\end{figure}

Figure~\ref{fig:fwm_6panel} shows the results of the 2-MOSD reconstruction (red) against the true distribution (black) and images resulting from confocal detection with (light blue) and without (dark blue) nuisance for the three models of FWM (figure columns) as well as four different scanning configurations (figure rows). Since global translations are of little significance in this study, the linear spectral phase of the distributions is matched to that of the true distribution. We also normalize the amplitude of all distributions to one. From there, the relative 2-norm errors are evaluated between the true distribution and distributions associated with 2-MOSD reconstruction ($\epsilon_\mathrm{2M}$), confocal detection with nuisance ($\epsilon_\mathrm{CN}$), and confocal detection without nuisance ($\epsilon_\mathrm{C}$). Those errors are displayed in the upper-left corners of the figure panels. The noisy $25\times25$ 2-MOSD reconstructions and $25\times1$ confocal images were generated with equal total acquisition times but differing integration times of $T$ and $25T$ per gate-position pair, respectively. Each noisy trace shown in the figure corresponds to the Poisson-noise realization whose relative 2-norm error was nearest the median of 1000 independent realizations (see section 2E of the SI).

In the absence of shot noise, the 2-MOSD reconstruction errors for models 1 through 3 are \((0.041,\,0.294,\,0.432)\) for the \(193\times193\) scan and \((0.013,\,0.287,\,0.457)\) for the \(25\times25\) scan. The similarity between these two sets of errors indicates that reconstruction accuracy is primarily limited by departures of the measurement from a 2-MOSD rather than scan-window size or sampling density. Introducing shot noise to the \(25\times25\) scan, the corresponding errors become \((0.186,\,0.318,\,0.458)\) for \(T=100~\mathrm{ms}\) and \((0.496,\,0.548,\,0.466)\) for \(T=10~\mathrm{ms}\), and this degradation is expected as the number of detected photons is reduced. Nevertheless, across all three FWM models and all four acquisition conditions, the 2-MOSD reconstruction has lower error than the confocal images with and without nuisance acquired at the same total acquisition time. In section 2F of the SI, we further present the extracted nuisance and gate transfer functions along with resolution comparisons between the 2-MOSD reconstruction and confocal detection.

\section{Outlook}

Considering the generality of structured detection and its associated benefits, future work should explore implementations of N-MOSDs beyond the spatial domain. In this paper, we numerically showed how FWM can be used to apply a 2-MOSD to an EMD in the time domain. Similar schemes based on FWM \cite{Janszky}, third-harmonic generation \cite{Janszky,Liu}, and cascaded second-order nonlinear processes \cite{Paulter,Feurer} have been proposed or implemented experimentally, but only for performing temporal third-order intensity autocorrelations. More relevantly, Wang et al. used cascaded Kerr gates to sample an EMD in time with two distinct pulses, although they did not attempt to redundantly encode the EMD \cite{Wang}. Expanding on this, the doubly gated setup could be preceded by frequency-to-time mapping to apply a 2-MOSD to an EMD in the optical-frequency domain; Fenwick et al. already demonstrated a single-gate implementation of such an experiment \cite{Fenwick}. In the radio-frequency domain, Joseph et al. modeled measurement with a spectrum analyzer as the convolution of the signal power spectrum with a shift-invariant filter \cite{Joseph}, and applying another shift-invariant filter with an independently tunable center frequency before detection with the analyzer could yield a 2-MOSD. Outside of EMDs, Cladé et al. laid much of the theoretical and experimental groundwork for applying a 2-MOSD to atomic number-density distributions in velocity space using velocity-selective Raman transitions \cite{Clade}. Indeed, restoring a nonuniform input distribution to their equation (19) gives the form of a 2-MOSD since the Raman resonance conditions are approximately affine in velocity. That being said, their experiment varied only one of the Raman detunings, so the measurement apparatus was equivalent to a 1-MOSD. We expect that other measurement schemes in the literature can similarly be interpreted as N-MOSDs and adapted to expand the frequency support through the introduction of additional kernels or, when the kernel centers can be varied independently, to exploit redundant encoding.

\section{Conclusion}

To conclude, we mathematically show that sampling an arbitrary distribution with multiple translated kernels can expand its accessible frequency support and, when the kernel centers are independently tunable, redundantly encode the distribution. The latter can help with extracting nuisance signals, estimating unknown kernels, and potentially improving resolution. We further support these findings with numerical studies showing how FWM can be used to implement structured detection of an EMD in the time domain. The resulting redundant encoding enables nuisance removal and gate-profile extraction for deconvolving the image, improving reconstruction accuracy relative to the temporal analogue of confocal detection even in the absence of nuisance. Together, these results show that the benefits of structured detection are not restricted to spatial imaging: they follow from the measurement structure itself and may therefore be transferred to time, frequency, velocity space, and other physical domains.

% ============================================================
% SUPPLEMENTAL DOCUMENT
% ============================================================
\clearpage
\nolinenumbers

% Restart SI section/equation/figure/table numbering while keeping
% internal PDF hyperlink anchors unique.
\setcounter{section}{0}
\setcounter{subsection}{0}
\setcounter{equation}{0}
\setcounter{figure}{0}
\setcounter{table}{0}

\renewcommand{\thesection}{\arabic{section}}
\renewcommand{\thesubsection}{\Alph{subsection}}
\renewcommand{\theequation}{S\arabic{equation}}
\renewcommand{\thefigure}{S\arabic{figure}}
\renewcommand{\thetable}{S\arabic{table}}

% Avoid duplicate hyperlink destinations after restarting counters.
\providecommand{\theHsection}{}
\providecommand{\theHsubsection}{}
\providecommand{\theHequation}{}
\providecommand{\theHfigure}{}
\providecommand{\theHtable}{}
\renewcommand{\theHsection}{SI.\arabic{section}}
\renewcommand{\theHsubsection}{SI.\arabic{section}.\Alph{subsection}}
\renewcommand{\theHequation}{SI.\arabic{equation}}
\renewcommand{\theHfigure}{SI.\arabic{figure}}
\renewcommand{\theHtable}{SI.\arabic{table}}

\begin{center}
{\Large\bfseries Structured detection beyond spatial imaging: supplemental document}
\end{center}
\vspace{1em}

\section{Applications of Redundant Encoding}

\subsection{Nuisance extraction}

For a fixed distribution frequency $q$, define the hyperplane
\begin{equation}
\mathcal{S}_q = \left\{ \bm{k}: \sum_{i=0}^{N-1}k_i=q \right\}, \qquad \bm{k}=\begin{bmatrix} k_0&\ldots &k_{N-1}\end{bmatrix}^T.
\end{equation}
Through the N-MOSD, every point $\bm{k}\in\mathcal{S}_q$ is associated with the same distribution coefficient $\tilde{\rho}(q)$ but with generally different encoding
\begin{equation}
A(\bm{k}) = \prod_{i=0}^{N-1}\tilde{h}_i(-k_i).
\end{equation}
Therefore, the data associated with N-MOSDs having tunable kernel centers ('general data') include multiple data points of the form
\begin{equation}
D(\bm{k}) = A(\bm{k})\tilde{\rho}(q), \qquad \bm{k}\in\mathcal{S}_q.
\end{equation}
Suppose instead that $L$ data points at $\bm{k}^{(\ell)}\in\mathcal{S}_q$, each labeled by $\ell$, are measured with a systematic contribution from an unknown nuisance signal $\tilde{B}$, whose values over the $L$ measurements are unrestricted, and a random zero-mean contribution from noise $n_\ell$:
\begin{equation}
Y_\ell(q) = A\left(\bm{k}^{(\ell)}\right)\tilde{\rho}(q) + \tilde{B}\left(\bm{k}^{(\ell)}\right) + n_\ell(q).
\end{equation}
Collecting the experimental data on $\mathcal{S}_q$ produces
\begin{equation}
\bm{Y}_q = \bm{a}_q\tilde{\rho}(q) + \bm{B}_q + \bm{n}_q,
\end{equation}
where
\begin{equation}
\bm{a}_q =
\begin{bmatrix}
A\left(\bm{k}^{(1)}\right) & \cdots & A\left(\bm{k}^{(L)}\right)
\end{bmatrix}^T,
\qquad
\bm{B}_q =
\begin{bmatrix}
\tilde{B}\left(\bm{k}^{(1)}\right) & \cdots & \tilde{B}\left(\bm{k}^{(L)}\right)
\end{bmatrix}^T.
\end{equation}
Since $\bm{Y}_q$ constitutes $L$ equations with $L+1$ unknowns, $\bm{B}_q$ cannot be determined from the experimental data alone. That being said, writing the nuisance vector as
\begin{equation}
\bm{B}_q = \alpha(q)\bm{a}_q + \bm{B}_q^\perp, \qquad \bm{a}_q^\dagger\bm{B}_q^\perp=0,
\label{eq:a}
\end{equation}
allows the measurements to be expressed as follows:
\begin{equation}
\bm{Y}_q = \bm{a}_q \left[ \tilde{\rho}(q)+\alpha(q) \right] + \bm{B}_q^\perp + \bm{n}_q.
\label{eq:b}
\end{equation}
Now, if $\tilde{\rho}(q) + \alpha(q)$ is considered a single unknown, equations~(\ref{eq:a}) and~(\ref{eq:b}) provide an equal number of equations and unknowns, and therefore $\bm{B}_q^\perp$ can be estimated. To isolate $\bm{B}_q^\perp$, apply the projection operator
\begin{equation}
\bm{P}_{\perp,q} = \bm{I} - \frac{ \bm{a}_q\bm{a}_q^\dagger }{ \bm{a}_q^\dagger\bm{a}_q}
\end{equation}
to equation~(\ref{eq:b}):
\begin{equation}
\bm{P}_{\perp,q}\bm{Y}_q = \bm{B}_q^\perp + \bm{P}_{\perp,q}\bm{n}_q.
\end{equation}
The least-squares solution for the perpendicular nuisance is
\begin{equation}
\hat{\bm{B}}_q^\perp = \bm{P}_{\perp,q}\bm{Y}_q.
\end{equation}
This estimate can then be subtracted from $\bm{Y}_q$ to, at minimum, partially reject nuisance, and since this protocol is not based on inversion, it does not amplify measurement noise. 

More complete nuisance extraction becomes possible when some structure of the nuisance is known. For instance, in some systems, the nuisance depends on fewer measurement coordinates than the desired signal. A nuisance may depend only on the center of one kernel, such that
\begin{equation}
B(\lambda_0,\ldots,\lambda_{N-1})=b_i(\lambda_i),
\end{equation}
or only on the relative position of two kernels, giving
\begin{equation}
B(\lambda_0,\ldots,\lambda_{N-1}) = b_{ij}(\lambda_i-\lambda_j).
\end{equation}
Such restrictions limit $B$ to a lower dimension of $k$-space and may consequently confine $\bm{B}_q$ to a lower-dimensional subspace 
\begin{equation}
\mathcal{U}_q\subset\mathbb{C}^L, \qquad \operatorname{dim}(\mathcal{U}_q)=P<L.
\end{equation}
The nuisance in this lower-dimensional subspace can be then expressed as
\begin{equation}
\bm{B}_q = \bm{U}_q\bm{\beta}_q,
\end{equation}
where $\bm{\beta}_q\in\mathbb{C}^P$ contains the unknown nuisance coefficients and the columns of $\bm{U}_q\in\mathbb{C}^{L\times P}$, determined by estimating which data points will be affected by the $P$ nuisances or perhaps guessed as part of a more advanced optimization algorithm, form a basis for $\mathcal{U}_q$. The measurement model then becomes
\begin{equation}
\bm{Y}_q = \bm{a}_q\tilde{\rho}(q) + \bm{U}_q\bm{\beta}_q + \bm{n}_q,
\end{equation}
and the unknowns can be solved for using least squares:
\begin{equation}
\hat{\bm{x}}_q=\underset{\bm{x}}{\operatorname{argmin}} \left\| \bm{Y}_q-\bm{H}_q\bm{x} \right\|^2_2=\left( \bm{H}_q^\dagger\bm{H}_q\right)^{-1}\bm{H}^\dagger_q\bm{Y}_q,
\end{equation}
where
\begin{equation}
\bm{H}_q = \begin{bmatrix} \bm{a}_q & \bm{U}_q \end{bmatrix}, \qquad \bm{x}=\begin{bmatrix} \tilde{\rho}(q) & \bm{\beta}_q \end{bmatrix}^T.
\end{equation}
A unique solution requires
\begin{equation}
\operatorname{rank} \begin{bmatrix}\bm{a}_q & \bm{U}_q \end{bmatrix}=P+1,
\label{rank}
\end{equation}
or, equivalently, $\bm{a}_q\notin\mathcal{U}_q$. This rank requirement may be of little physical relevance for nuisance generated inside the measurement apparatus. However, if the nuisance originates before the apparatus and couples into the distribution mode, then this requirement will not be met.

\subsection{Kernel extraction}

Let the experimentally measured data be modeled as
\begin{equation}
Y(\bm{k}) = \tilde{\rho}(q)\prod_{i=0}^{N-1}\tilde{h}_i(-k_i) + n(\bm{k})
\end{equation}
Now suppose estimates of the kernels $\tilde{h}_i^{(r)}$ are available at iteration $r$. Since every measurement at $\bm{k}\in S_q$ contains the same $\tilde{\rho}$, the distribution coefficient can be estimated using linear least-squares:
\begin{equation}
\hat{\tilde{\rho}}^{(r+1)}(q) = \frac{ \displaystyle\sum_{\bm{k}\in S_q} [A^{(r)}(\bm{k})]^*Y(\bm{k}) }{ \displaystyle\sum_{\bm{k}\in S_q} |A^{(r)}(\bm{k})|^2 }, \qquad
A^{(r)}(\bm{k}) = \prod_{i=0}^{N-1}\hat{\tilde{h}}_i^{(r)}(-k_i).
\end{equation}
Similarly, for a fixed kernel frequency $\kappa=k_i$, all of the measurements on the corresponding hyperplane $S_{i,\kappa}=\left\{ \bm{k}:k_i=\kappa \right\}$ satisfy
\begin{equation}
Y(\bm{k}) = \tilde{h}_i(-\kappa)C_i^{(r)}(\bm{k})+n(\bm{k}),
\qquad C_i^{(r)}(\bm{k}) = \hat{\tilde{\rho}}^{(r+1)} \left(\sum_{j=0}^{N-1}k_j\right) \prod_{\substack{j=0\\j\neq i}}^{N-1} \hat{\tilde{h}}_j^{(r)}(-k_j).
\end{equation}
The least-squares update is therefore
\begin{equation}
\hat{\tilde{h}}_i^{(r+1)}(-\kappa) = \frac{ \displaystyle\sum_{\bm{k}\in S_{i,\kappa}} [C_i^{(r)}(\bm{k})]^*Y(\bm{k}) }{ \displaystyle\sum_{\bm{k}\in S_{i,\kappa}} |C_i^{(r)}(\bm{k})|^2 }.
\label{kernel}
\end{equation}
After updating the kernels, the distribution can be estimated again and the procedure repeated until convergence. It is worth noting that $\hat{\tilde{\rho}}$ and $\hat{\tilde{h}}_i$ become sensitive to noise when $(\sum_{\bm{k}\in S_q}|A^{(r)}(\bm{k})|^2)^{-1}$ and $(\sum_{\bm{k}:k_i=\kappa}|C_i^{(r)}(\bm{k})|^2)^{-1}$ are large, respectively. 
More advanced algorithms can incorporate additional model constraints or statistical information to stabilize the inversion \cite{SI-Zunino2023,SI-Lim2026}. Ambiguity in the scaling of the kernels and distribution must also be addressed, and this can be done by enforcing $\hat{\tilde{h}}_i(0)=1$ for integral-normalized kernels for each update. Similarly, letting $\hat{\tilde{\rho}}(q)\rightarrow e^{cq}\hat{\tilde{\rho}}(q)$ and $\hat{\tilde{h}}_i(-k)\rightarrow e^{-ck} \hat{\tilde{h}}_i(-k)$ for a complex constant $c=c_R+ic_I$ also reproduces the same $\hat{Y}(\bm{k})$. A purely imaginary $c$ corresponds to a translation of all the kernels and distribution by the same amount, which is often harmless. On the other hand, a real $c$ corresponds to an exponential spectral tilt and can be rather problematic. However, if the distribution or one of the kernels is known to be real, $c_R=0$ by Hermitian symmetry of their transfer functions. Therefore, enforcing a real kernel or distribution by taking the real part of the inverse Fourier transform of its estimate and Fourier transforming again during each update can remove the tilt ambiguity.

\subsection{Preventing attenuation of total transfer function}

Data associated with N-MOSDs having fixed relative kernel centers ('diagonal data') are of the form
\begin{equation}
D_{\mathrm{diag}}(q) = \tilde{\rho}(q)H_{\mathrm{diag}}(q) + n(q), \qquad
H_{\mathrm{diag}}(q) = \frac{1}{(2\pi)^{N-1}}\int_{S_q} A(\bm{k})\,d^{N-1}\bm{k},
\end{equation}
so if $A(\bm{k})$ varies in sign or phase over $\mathcal{S}_q$, the measurement can partially or completely vanish despite a finite $\tilde{\rho}(q)$. In contrast, general N-MOSD data over the same $\mathcal{S}_q$ are
\begin{equation}
D_{\mathrm{gen}}(\bm{k}) = A(\bm{k})\tilde{\rho}(q) + n(\bm{k}), \qquad \bm{k}\in S_q.
\label{eq:c}
\end{equation}
From this, a data set resembling $D_\mathrm{diag}$ can be synthesized using a weighting function $W_q(\bm{k})$:
\begin{equation}
D^\mathrm{syn}_\mathrm{gen}(q) = \frac{1}{(2\pi)^{N-1}}\int_{S_q} W_q(\bm{k}) D_{\mathrm{gen}}(\bm{k}) \,d^{N-1}\bm{k}.
\label{eq:d}
\end{equation}
Substituting equation~(\ref{eq:c}) into~(\ref{eq:d}) then gives
\begin{equation}
D^\mathrm{syn}_\mathrm{gen}(q) = \tilde{\rho}(q)H^\mathrm{syn}_\mathrm{gen}(q) + n^\mathrm{syn}_\mathrm{gen}(q),
\end{equation}
where
\begin{equation}
H^\mathrm{syn}_\mathrm{gen}(q) = \frac{1}{(2\pi)^{N-1}}\int_{S_q} W_q(\bm{k})A(\bm{k}) \,d^{N-1}\bm{k}
\end{equation}
and
\begin{equation}
n^\mathrm{syn}_\mathrm{gen}(q) = \frac{1}{(2\pi)^{N-1}}\int_{S_q} W_q(\bm{k})n(\bm{k}) \,d^{N-1}\bm{k}.
\end{equation}
Choosing
\begin{equation}
W_q(\bm{k}) = e^{-i\arg[A(\bm{k})]}
\end{equation}
then gives
\begin{equation}
H^\mathrm{syn}_\mathrm{gen}(q) = \frac{1}{(2\pi)^{N-1}}\int_{S_q} |A(\bm{k})| \,d^{N-1}\bm{k}.
\end{equation}
In words, general N-MOSD data can be manipulated to reproduce diagonal N-MOSD data but with the individual kernels' transfer functions being replaced by their magnitudes. Since
\begin{equation}
\left|\int_{S_q} A(\bm{k})\,d^{N-1}\bm{k}\right| \leq \int_{S_q} |A(\bm{k})| \,d^{N-1}\bm{k},
\end{equation}
this can help prevent attenuation of the overall transfer function of the combined system ('total transfer function') associated with complex or mixed-sign transfer functions of individual kernels. 

\subsection{Resolution improvement}

One class of weighting functions that has been frequently used in the past to improve resolution over diagonal data selects a single encoding of each distribution frequency
\begin{equation}
W_q^{\star}(\bm{k}) = (2\pi)^{N-1} \prod_{i=1}^{N-1} \delta\!\left[k_i-k_i^{\star}(q)\right], \qquad \bm{k}^{\star}(q) = \begin{bmatrix} k_0^{\star}(q) & \cdots & k_{N-1}^{\star}(q) \end{bmatrix}^{T} \in S_q,
\end{equation}
yielding a transfer function of the form
\begin{equation}
H_{\star}(q) = A\!\left(\bm{k}^{\star}(q)\right) = \prod_{i=0}^{N-1}\tilde{h}_i\!\left(-k_i^{\star}(q)\right).
\end{equation}
For instance, in ISM \cite{SI-Sheppard1988,SI-Muller2010}, the weight is set to 
\begin{equation}
W_q^{\mathrm{ISM}}(\kappa) = 2\pi\delta\left(\kappa-\frac{q}{2}\right)
\end{equation}
so that the total transfer function associated with the 2-MOSD data becomes
\begin{equation}
H_{\mathrm{ISM}}(q) = \tilde{h}_0\left(-\frac{q}{2}\right) \tilde{h}_1\left(-\frac{q}{2}\right).
\end{equation}
Similarly, in VSD \cite{SI-Lu2013}, a sinusoidal pattern of frequency $K$ applied to the distribution digitally can, after appropriate manipulation of the measured data, effectively produce weights
\begin{equation}
W_q^{\mathrm{VSD\pm}}(\kappa) = 2\pi \delta(\kappa\pm K)
\end{equation}
to give the total transfer functions
\begin{equation}
H_{\mathrm{VSD\pm}}(q) = \tilde{h}_0(-(q\mp K))\tilde{h}_1(\mp K).
\end{equation}
Whether $H_{VSD+}$ or $H_{VSD-}$ is chosen often depends on the sign of q.  

A simple example demonstrates how these weights alter resolution. Suppose the two kernels are identical $\mathrm{sinc}^2$ functions so that their transfer functions are
\begin{equation}
\tilde{h}_0(-k)=\tilde{h}_1(-k) = \begin{cases} 1-|k|/k_c, & |k|\leq k_c,\\ 0, & |k|>k_c, \end{cases}
\end{equation}
where $k_c$ denotes the cutoff frequency. After normalization at $q=0$, the diagonal total transfer function in the range $k_c\leq q\leq2k_c$ is
\begin{equation}
\mathcal{H}_{\mathrm{diag}}(q) = \frac{H_\mathrm{diag}(q)}{H_\mathrm{diag}(0)}=\frac{1}{4} \left( 2-\frac{q}{k_c} \right)^3.
\end{equation}
On the other hand, the ISM transfer function is
\begin{equation}
H_{\mathrm{ISM}}(q) = \left( 1-\frac{q}{2k_c} \right)^2 = \frac{1}{4} \left( 2-\frac{q}{k_c} \right)^2,
\end{equation}
and, after setting $K=q/2$ to maximize the transfer function at any given $q$, VSD also has
\begin{equation}
H_{\mathrm{VSD,opt}}^{+}(q) = \frac{1}{4} \left( 2-\frac{q}{k_c} \right)^2, \qquad 0\leq K\leq k_c, \qquad 0\leq q-K\leq k_c.
\end{equation}
For these normalized transfer functions,
\begin{equation}
\frac{H_{\mathrm{ISM,VSD}}(q)} {\mathcal{H}_{\mathrm{diag}}(q)} = \frac{1} {2-q/k_c}, \qquad k_c<q<2k_c.
\end{equation}
Thus, even though diagonal and general data are restrained by the same frequency support, a suitable weighting of general N-MOSD data can make the total transfer function near the cutoff behave more favorably than the total transfer function associated with diagonal data. Whether this improved behavior near the cutoff translates into a higher SNR for high distribution frequencies ultimately depends on the specifics of the systems used to acquire the diagonal and general data.

\section{Temporal 2-MOSD Simulations}

\subsection{Models of FWM}

Following the treatment of FWM by Agrawal \cite{SI-Agrawal}, consider four copolarized fields with carrier frequencies $\omega_j$ satisfying $\omega_1+\omega_2=\omega_3+\omega_4$, where subscripts $1$ through $4$ respectively denote the first gate, second gate, distribution, and idler. Assuming the fields have approximately the same transverse mode profiles that are invariant with propagation, the equations governing the evolution of the slowly-varying envelopes $A_j$ are
\begin{align}
\left(\frac{\partial}{\partial z} -\Delta\beta_{1,1}\frac{\partial}{\partial \tau} +i\frac{\beta_{2,1}}{2}\frac{\partial^2}{\partial \tau^2}\right)A_1
&= \frac{in_2\omega_1}{cA_{\mathrm{eff}}} \left[ \left(|A_1|^2+2\sum_{m\neq1}|A_m|^2\right)A_1 +2A_2^*A_3A_4e^{i\Delta kz} \right],\\ \left(\frac{\partial}{\partial z} -\Delta\beta_{1,2}\frac{\partial}{\partial \tau} +i\frac{\beta_{2,2}}{2}\frac{\partial^2}{\partial \tau^2}\right)A_2
&= \frac{in_2\omega_2}{cA_{\mathrm{eff}}} \left[ \left(|A_2|^2+2\sum_{m\neq2}|A_m|^2\right)A_2 +2A_1^*A_3A_4e^{i\Delta kz} \right],\\ \left(\frac{\partial}{\partial z} -\Delta\beta_{1,3}\frac{\partial}{\partial \tau} +i\frac{\beta_{2,3}}{2}\frac{\partial^2}{\partial \tau^2}\right)A_3 &= \frac{in_2\omega_3}{cA_{\mathrm{eff}}} \left[ \left(|A_3|^2+2\sum_{m\neq3}|A_m|^2\right)A_3 +2A_4^*A_1A_2e^{-i\Delta kz} \right],\\ \left(\frac{\partial}{\partial z} +i\frac{\beta_{2,4}}{2}\frac{\partial^2}{\partial \tau^2}\right)A_4
&= \frac{in_2\omega_4}{cA_{\mathrm{eff}}} \left[ \left(|A_4|^2+2\sum_{m\neq4}|A_m|^2\right)A_4 +2A_3^*A_1A_2e^{-i\Delta kz} \right],
\end{align}
where $n_2$ is the nonlinear index; $A_\mathrm{eff}$ is the effective mode area; $c$ is the speed of light; the mismatch between the wavenumbers $k_j$ is
\begin{equation}
\Delta k=k_3+k_4-k_1-k_2;
\end{equation}
the first- and second-order dispersion coefficients are, respectively,
\begin{equation}
\beta_{1,j} = \left.\frac{dk}{d\omega}\right|_{\omega_j} \quad \mathrm{and} \quad \beta_{2,j} = \left.\frac{d^2k}{d\omega^2}\right|_{\omega_j};
\end{equation}
\begin{equation}
\Delta\beta_{1,j}=\beta_{1,4}-\beta_{1,j};
\end{equation}
and the equations are written in the retarded time $\tau=t-\beta_{1,4}z$ corresponding to a frame moving with the idler group velocity. The equations above describe model 3, and, in general, they do not admit an analytic solution.

To obtain model 2, set $\Delta k=0$ and $\beta_{2,j}=0$, neglect SPM and XPM, and assume negligible depletion of the three input fields such that
\begin{equation}
\left( \frac{\partial}{\partial z} -\Delta\beta_{1,j}\frac{\partial}{\partial \tau} \right)A_j=0, \qquad j=1,2,3
\label{simp}
\end{equation}
while retaining the FWM source term in the propagation equation for the idler. Using the initial conditions
\begin{equation}
A_1(0,\tau;t_1)=a_1(\tau-t_1),\qquad A_2(0,\tau;t_2)=a_2(\tau-t_2),\qquad A_3(0,\tau)=a_3(\tau),
\end{equation}
where $t_1$ and $t_2$ denote the delays of the two gates relative to the distribution, the corresponding solutions to equation~(\ref{simp}) are
\begin{equation}
A_j(z,\tau;t_j) = a_j(\tau+\Delta\beta_{1,j}z-t_j),
\end{equation}
assuming $t_3=0$. The idler propagation equation then becomes
\begin{equation}
\frac{\partial A_4}{\partial z} = \frac{2in_2\omega_4}{cA_{\mathrm{eff}}} a_1(\tau+\Delta\beta_{1,1}z-t_1) a_2(\tau+\Delta\beta_{1,2}z-t_2) a_3^*(\tau+\Delta\beta_{1,3}z).
\end{equation}
For $A_4(0,\tau)=0$, integration over a nonlinear-medium length $L$ gives
\begin{equation}
A_4(L,\tau;t_1,t_2) = \frac{2in_2\omega_4}{cA_{\mathrm{eff}}} \int_0^L a_1(\tau+\Delta\beta_{1,1}z-t_1) a_2(\tau+\Delta\beta_{1,2}z-t_2) a_3^*(\tau+\Delta\beta_{1,3}z) \,dz,
\label{18}
\end{equation}
and integrating the intensity over time then yields
\begin{equation}
D_2(t_1,t_2) \propto \int \left| \int_0^L a_1(\tau+\Delta\beta_{1,1}z-t_1) a_2(\tau+\Delta\beta_{1,2}z-t_2) a_3^*(\tau+\Delta\beta_{1,3}z) \,dz \right|^2d\tau.
\label{19}
\end{equation}
As can be seen, the ideal 2-MOSD is recovered when $\Delta\beta_{1,j}=0$, so model 2 provides a departure from the 2-MOSD governed by $\Delta\beta_{1,j}$.

As suggested above, model 1 is obtained from model 2 by setting $\Delta\beta_{1,j} = 0$. Equation~(\ref{18}) then reduces to
\begin{equation}
A_4(L,\tau;t_1,t_2) = \frac{2in_2\omega_4L}{cA_{\mathrm{eff}}} a_1(\tau-t_1)a_2(\tau-t_2)a_3^*(\tau),
\end{equation}
and integrated detection of the idler intensity gives
\begin{equation}
D_1(t_1,t_2) \propto \int |a_1(\tau-t_1)|^2 |a_2(\tau-t_2)|^2 |a_3(\tau)|^2d\tau.
\end{equation}
This is exactly the form of a 2-MOSD.

\subsection{Additional waveguide parameters}

The bulk indices for stoichiometric $\mathrm{Si_3N_4}$ and $\mathrm{SiO_2}$ given by Luke \cite{SI-Luke} and Malitson \cite{SI-Malitson}, respectively, are used in solving the waveguide modes. The effective area is approximated from the mode fields of the waveguide as
\begin{equation}
\frac{1}{A_{\mathrm{eff}}} = \frac{ \left| \displaystyle\int_{\mathrm{Si_3N_4}} E_1 E_2 E_3^* E_4^*\,dA \right| }{ \displaystyle \sqrt{ \prod_{j=1}^{4}
\int |E_j|^2\,dA } },
\end{equation}
where $E_j$ denotes the electric-field component of the $j$th mode along the TE polarization direction, yielding $A_{\mathrm{eff}}=0.775~\mu\mathrm{m}^2$. The nonlinear index of $\mathrm{Si_3N_4}$ is set to $2.4\times10^{-19}\ \mathrm{m^2/W}$ \cite{SI-Ikeda}.

\subsection{Additional simulation parameters}

The input envelopes described in the main text are initialized on a $8000\ \mathrm{fs}$ grid with $1.95\ \mathrm{fs}$ spacing that comfortably handles the relative delays and pulse shapes involved in the simulation. Based on a relative Frobenius-norm error of $\sim10^{-5}$ with respect to a reference 2-MOSD dataset produced with a $0.05\ \mathrm{mm}$ step size, the propagation step size is set to $0.1\ \mathrm{mm}$.

\subsection{Reconstruction algorithm}

The nuisance along with the transfer functions of the gates are estimated in an iterative algorithm to recover the distribution. More specifically, at the first iteration, the transfer functions are initialized as flat-top functions up to their cutoff frequencies, which are assumed to be known and have values of $2\pi/T_0$. These transfer functions are used to jointly solve for the distribution and nuisance, as described in the latter half of section 1A, in the first linear least squares (LLS) problem of the algorithm. To enact the rank condition of section 1A, we discard the $q$ values that have a reciprocal condition number of $\bar{\bm{H}}_q^\dagger\bar{\bm{H}}_q$ that is less than $10^{-6}$, where $\bar{\bm{H}}_q$ is $\bm{H}_q$ with normalized columns. The $q$ values for which $||\bm{a}_q||_2^2$ is less than $10^{-6}$ of its maximum value are also excluded from the extraction. Once the nuisance is estimated, it is subtracted from the measurement data. Following section 1B, those measurement data, the estimate of the distribution from the first LLS problem, and the initial or previous-iteration transfer function of the second gate are used to estimate the transfer function of the first gate in a second LLS problem; the second transfer function is solved for in a similar manner using the initial or previous-iteration transfer function of the first gate. Transfer-function frequencies for which the denominator in equation~(\ref{kernel}) is less than $10^{-6}$ of its maximum value are excluded from the corresponding transfer-function estimate. The transfer functions are then normalized to one at zero frequency and forced to obey Hermitian symmetry to avoid the ambiguities described in section 1B. Aside from those ambiguities, there is an additional ambiguity that must be considered for this specific iterative algorithm. Because the nuisance depends only on the first-gate delay, the Fourier-domain data can be written as
\begin{equation}
Y(k_1,k_2)
=
\tilde{\rho}(k_1+k_2)\tilde{h}_1(-k_1)\tilde{h}_2(-k_2)
+
\tilde{B}(k_1)\delta_{k_2,0},
\end{equation}
where $\delta_{k_2,0}$ is a Kronecker delta. Even after setting $\tilde{h}_2(0)=1$, the transformations
\begin{equation}
\tilde{\rho}(q)\rightarrow a\tilde{\rho}(q),
\qquad
\tilde{h}_2(k_2)\rightarrow a^{-1}\tilde{h}_2(k_2)
\label{61}
\end{equation}
leave the $Y$ at $k_2\neq0$ unchanged while the corresponding change at $k_2=0$ can be absorbed by
\begin{equation}
\tilde{B}(k_1)\rightarrow
\tilde{B}(k_1)
+
(1-a)\tilde{\rho}(k_1)\tilde{h}_1(-k_1).
\label{62}
\end{equation}
This additional ambiguity is addressed by approximating the first nonzero-frequency coefficient of the second-gate transfer function as unity since $\tilde{h}_2(\Delta k)\rightarrow\tilde{h}_2(0)=1$ as the Fourier-grid spacing $\Delta k\rightarrow0$ for a physical transfer function that is continuous near zero frequency. The corresponding negative-frequency coefficient is fixed by Hermitian symmetry. Once this additional constraint is imposed, the transfer functions are also clipped so that their magnitudes are less than or equal to $1$ throughout their domains. The joint distribution-and-nuisance solve, the first-gate transfer-function solve, and the second-gate transfer-function solve are then repeated until convergence. After convergence, the distribution and nuisance are estimated once more using the final transfer functions.

\subsection{Monte Carlo simulations}

For each measurement class, defined by the FWM model, modality, and finite integration time, 1000 images were generated using independent Poisson-noise realizations, and a relative 2-norm error was calculated for each image. The images shown in the bottom two rows of figure~1 were selected as realizations whose errors were nearest the median for their respective measurement classes. The resulting Monte Carlo statistics are summarized in table~\ref{tab:shot_noise}.

\begin{table}[h]
\centering
\caption{Relative 2-norm errors over 1000 independent Poisson-noise realizations. Values are mean $\pm$ standard deviation, with the median given in parentheses. $P_{\mathrm{2M<C}}$ denotes the fraction of realizations for which the 2-MOSD reconstruction error was lower than both confocal errors.}
\label{tab:shot_noise}
\begin{tabular}{cccccc}
\hline
$T$ & Model &
$\epsilon_{\mathrm{2M}}$ &
$\epsilon_{\mathrm{C}}$ &
$\epsilon_{\mathrm{CN}}$ &
$P_{\mathrm{2M<C}}$ \\
\hline
100 ms & 1 & $0.197\pm0.096\;(0.186)$ & $0.5472\pm0.0003\;(0.5472)$ & $0.6701\pm0.0002\;(0.6701)$ & $99.9\%$ \\
100 ms & 2 & $0.339\pm0.073\;(0.318)$ & $0.6579\pm0.0003\;(0.6578)$ & $0.6991\pm0.0002\;(0.6990)$ & $100\%$ \\
100 ms & 3 & $0.457\pm0.013\;(0.458)$ & $0.6819\pm0.0003\;(0.6819)$ & $0.7206\pm0.0001\;(0.7206)$ & $100\%$ \\
10 ms & 1 & $0.515\pm0.257\;(0.496)$ & $0.5469\pm0.0008\;(0.5469)$ & $0.6697\pm0.0006\;(0.6698)$ & $56.9\%$ \\
10 ms & 2 & $0.570\pm0.200\;(0.548)$ & $0.6574\pm0.0008\;(0.6574)$ & $0.6990\pm0.0006\;(0.6990)$ & $69.8\%$ \\
10 ms & 3 & $0.471\pm0.050\;(0.466)$ & $0.6819\pm0.0008\;(0.6819)$ & $0.7206\pm0.0004\;(0.7206)$ & $99.8\%$ \\
\hline
\end{tabular}
\end{table}

At $T=100~\mathrm{ms}$, the 2-MOSD reconstruction had lower error than the confocal images in essentially every realization. At $T=10~\mathrm{ms}$, this remained true for nearly all model-3 realizations but less consistently for models 1 and 2. The comparatively weak noise sensitivity of model 3 is consistent with its two outermost distribution-frequency coefficients falling below the common reconstruction threshold and being rejected during reconstruction, whereas the corresponding weakly encoded coefficients in models 1 and 2 remain just above threshold and are retained.

\subsection{Additional results}

\begin{figure}[h]
    \centering
    \includegraphics[width=\linewidth]{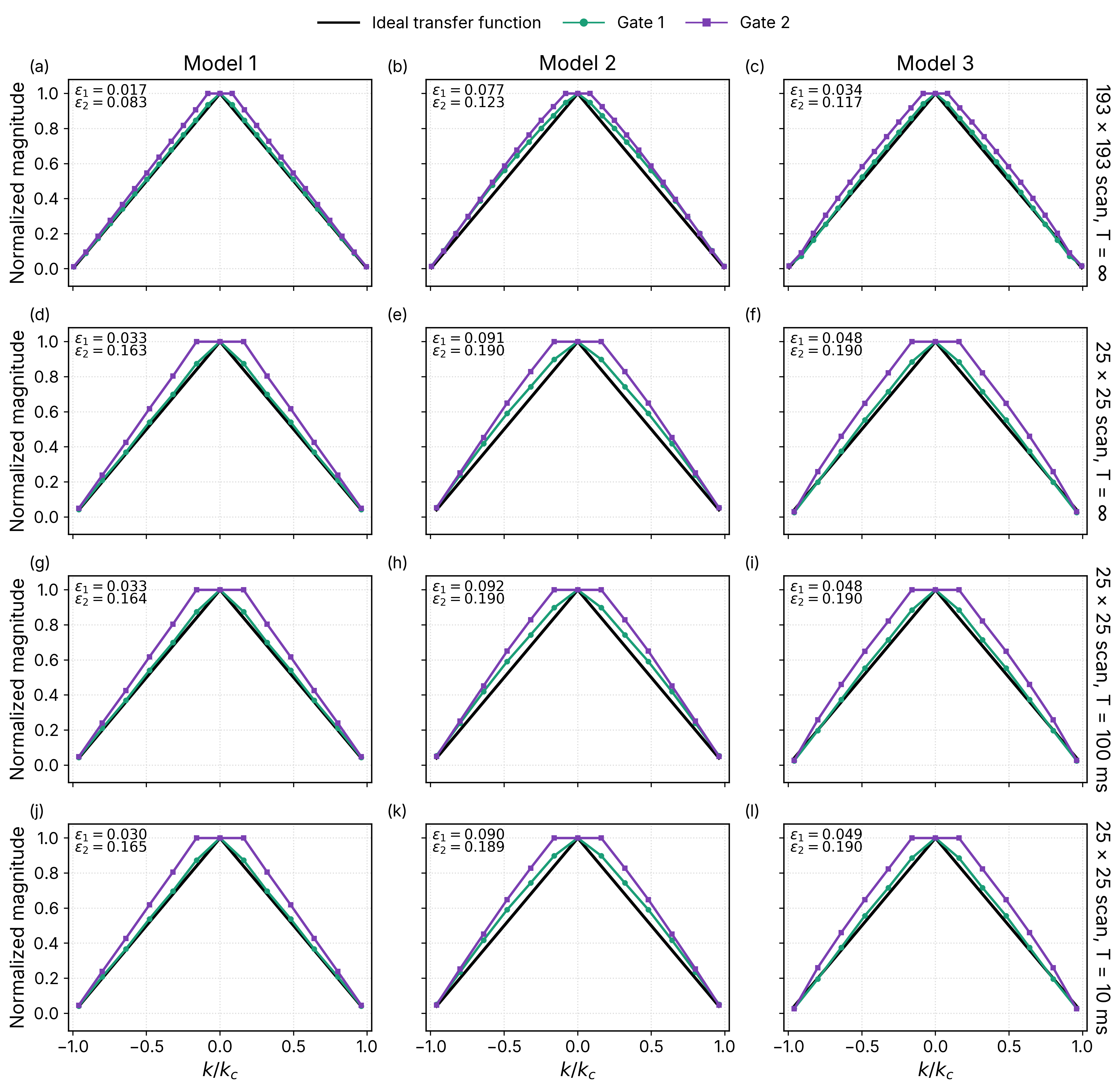}
    \caption{
    Extracted transfer functions for the three FWM models (columns) and four acquisition conditions (rows) corresponding to figure~1 of the main text. The ideal transfer function (black) is compared with the extracted transfer functions of the first (green) and second (purple) gates. The corresponding relative 2-norm errors $\epsilon_1$ and $\epsilon_2$ are given in the upper-left corner of each panel.
    }
    \label{fig:fwm_6panel_3}
\end{figure}

Figure~\ref{fig:fwm_6panel_3} shows the extracted transfer functions of the first (green) and second (purple) gates for the three FWM models (columns) and the four acquisition conditions considered in figure~1 of the main text (rows), with the ideal transfer function associated with the input gate profiles shown as well (black curve). For the $193\times193$ scan, the first-gate errors for models 1 through 3 are $(0.017,0.077,0.034)$, and the corresponding second-gate errors are $(0.083,0.123,0.117)$. For the noiseless $25\times25$ scan, the first- and second-gate errors are $(0.033,0.091,0.048)$ and $(0.163,0.190,0.190)$, respectively. Introducing shot noise produces only modest changes. For $T=100~\mathrm{ms}$, the corresponding errors are $(0.033,0.092,0.048)$ and $(0.164,0.190,0.190)$, while for $T=10~\mathrm{ms}$ they are $(0.030,0.090,0.049)$ and $(0.165,0.189,0.190)$.

The relatively large gate-2 transfer-function errors partly arise from the scale constraint imposed at the first nonzero frequency to remove the ambiguity defined by equations~\ref{61} and~\ref{62}. For an ideal sinc$^2$ gate, the first nonzero-frequency coefficient is $0.840$ for the $25\times25$ scan and $0.917$ for the $193\times193$ scan. Constraining these coefficients to unity therefore scales the remaining nonzero-frequency coefficients upward by factors of approximately $1.19$ and $1.09$ for the $25\times25$ and $193\times193$ scans, respectively. Applying the same scaling to the ideal transfer function produces relative errors of $0.166$ and $0.085$, closely matching the corresponding model-1 gate-2 errors of approximately $0.16$ and $0.08$. Thus, most of the gate-2 error in model 1 can be attributed to the imposed scale constraint. The somewhat larger gate-2 errors for models 2 and 3 are consistent with additional deviations introduced as the measurement departs from a 2-MOSD.

\begin{figure}[h]
    \centering
    \includegraphics[width=\linewidth]{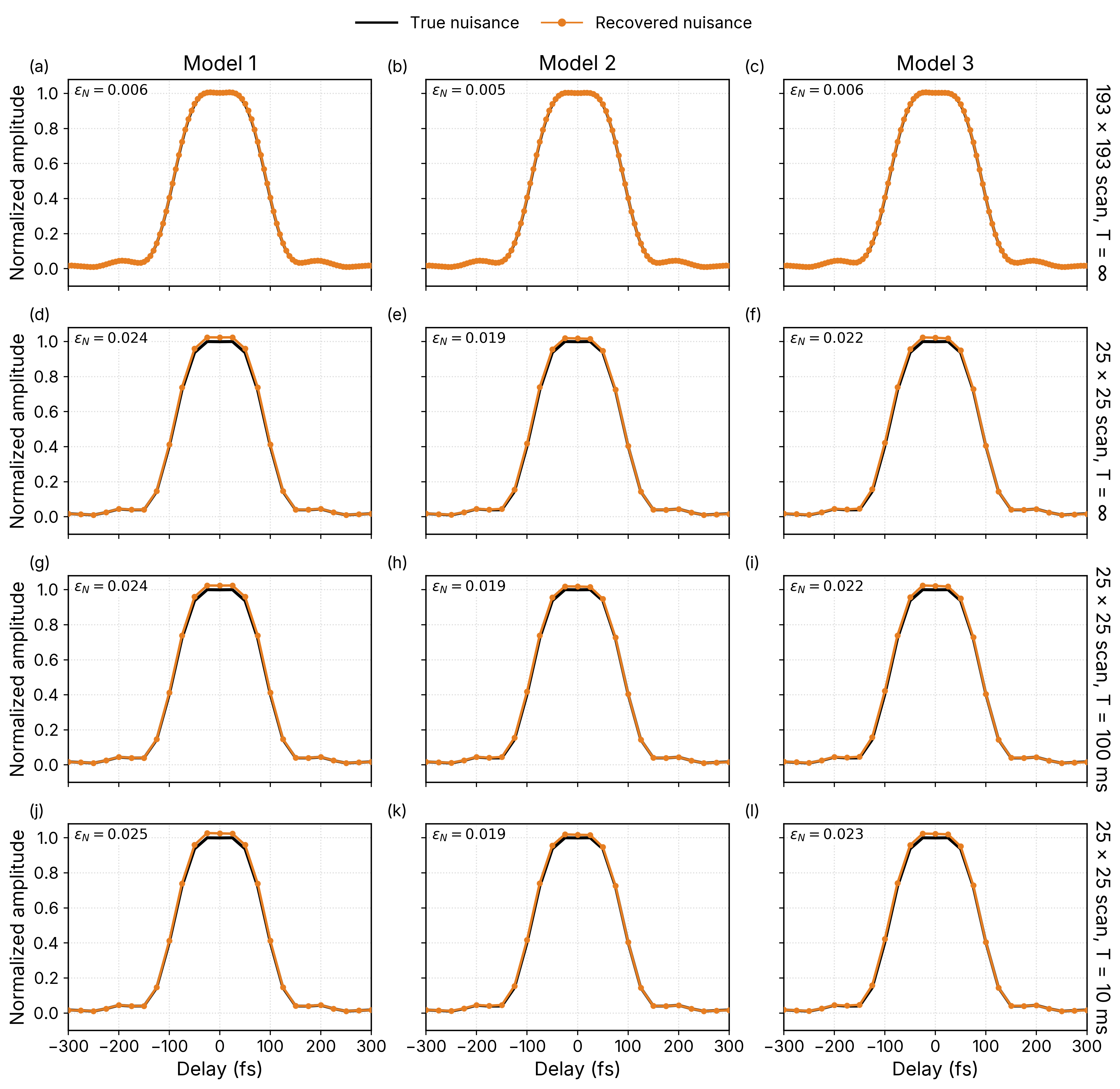}
    \caption{
    Extracted nuisance for the three FWM models (columns) and four acquisition conditions (rows) corresponding to figure~1 of the main text. The true nuisance signal (black) is compared with the recovered nuisance (orange). The corresponding relative 2-norm error $\epsilon_N$ is given in the upper-left corner of each panel.
    }
    \label{fig:fwm_6panel_2}
\end{figure}

Figure~\ref{fig:fwm_6panel_2} shows the true nuisance signal together with the nuisance extracted for the same four acquisition conditions. The corresponding errors for models 1 through 3 are $(0.006,0.005,0.006)$ for the $193\times193$ scan and $(0.024,0.019,0.022)$ for the noiseless $25\times25$ scan. The errors remain nearly unchanged after introducing shot noise, with values of $(0.024,0.019,0.022)$ for $T=100~\mathrm{ms}$ and $(0.025,0.019,0.023)$ for $T=10~\mathrm{ms}$. Because the recovered nuisance is not independently normalized before the error is evaluated, these errors include both shape and amplitude differences. The relatively weak dependence of the nuisance-extraction error on the FWM model and integration time indicates that the nuisance can still be separated accurately even as the measurement departs from a 2-MOSD and shot noise is introduced.

\begin{figure}[h]
    \centering
    \includegraphics[width=\linewidth]{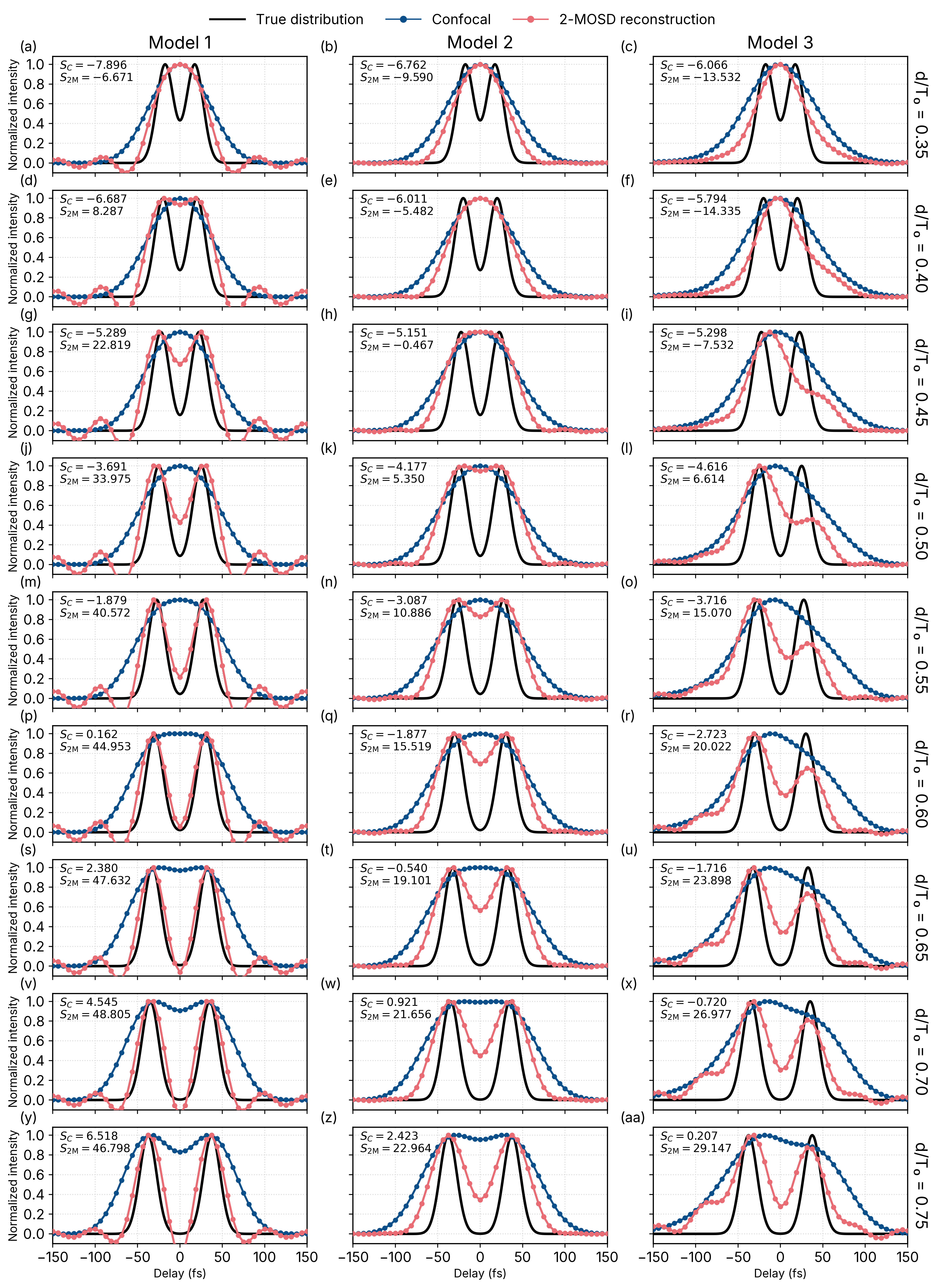}
    \caption{
    Full resolution comparison between confocal imaging and the 2-MOSD reconstruction for the three FWM models (columns) and normalized peak separations from $d/T_0=0.35$ to $0.75$ (rows). The true distribution (black), confocal image (blue), and 2-MOSD reconstruction (red) are shown after normalization and temporal alignment. The corresponding Sparrow parameters $S_C$ and $S_{\mathrm{2M}}$ are given in the upper-left corner of each panel.
    }
    \label{fig:fwm_27panel}
\end{figure}

Figure~\ref{fig:fwm_27panel} demonstrates the resolution advantage of the 2-MOSD reconstruction over confocal imaging in the slow-scan configuration. The normalized peak separation is varied from $d/T_0=0.35$ to $0.75$ (rows). The resolving power associated with the two modalities is quantified using a parameter derived from the Sparrow criterion \cite{SI-Sparrow1916}:
\begin{equation}
S_j=\left.
\frac{\partial^2 \bar{\rho}_j}{\partial t^2} \right|_{t=0}
\simeq
\frac{\bar{\rho}_j(\Delta t)
-2\bar{\rho}_j(0)
+\bar{\rho}_j(-\Delta t)}
{(\Delta t)^2}.
\end{equation}
These values are presented in the upper left-hand corners of the figure panels. For the purpose of this study, nuisance is removed, and the width of the Gaussians defining the distribution is also reduced. Ideally, the width would be allowed to approach zero, but reducing the width generally increases the Fourier coefficient of the distribution at twice the cutoff frequency of the gates, producing an increasingly sharp transition to zero for the 2-MOSD reconstructions at that frequency. In turn, this leads to pronounced ringing in the reconstructions. If the width is too large, on the other hand, images of the distribution cannot be faithfully used to estimate resolution. For instance, for a flat passband extending to $4\pi/(100\ \mathrm{fs})$, the Sparrow separation for Gaussian peaks with $\sigma=20\ \mathrm{fs}$ differs from the delta-function limit by approximately $30\%$, whereas the difference is only about $6\%$ for $\sigma=10\ \mathrm{fs}$. Accordingly, the peak width is set to $\sigma=10\ \mathrm{fs}$ as a compromise between the ringing and finite-width bias in the resolution estimate.

To estimate the Sparrow separation associated with each imaging modality, we linearly interpolate $S$ between the two nearest sampled separations for which the parameter changes sign and calculate the separation at which it crosses zero. For models 1 through 3, the resulting confocal Sparrow separations are $0.596T_0$, $0.668T_0$, and $0.739T_0$, respectively, whereas the corresponding 2-MOSD separations are $0.372T_0$, $0.454T_0$, and $0.477T_0$. Thus, compared to confocal imaging, the 2-MOSD reconstruction reduces the Sparrow separation by approximately $37.5\%$, $32.1\%$, and $35.5\%$ for the three models, respectively. These numbers should be interpreted with a degree of care, however. The model-3 reconstructions are notably asymmetric, whereas the Sparrow criterion was developed for symmetric distributions. In addition, ringing artifacts meaningfully affect the model-1 and model-3 reconstructions. For instance, at $d=0.75T_0$, the normalized spectral magnitudes of the reconstructions near twice the cutoff frequency are $0.214$, $0.024$, and $0.227$ for models 1 through 3, respectively.

\begin{figure}[h]
    \centering
    \includegraphics[width=\linewidth]{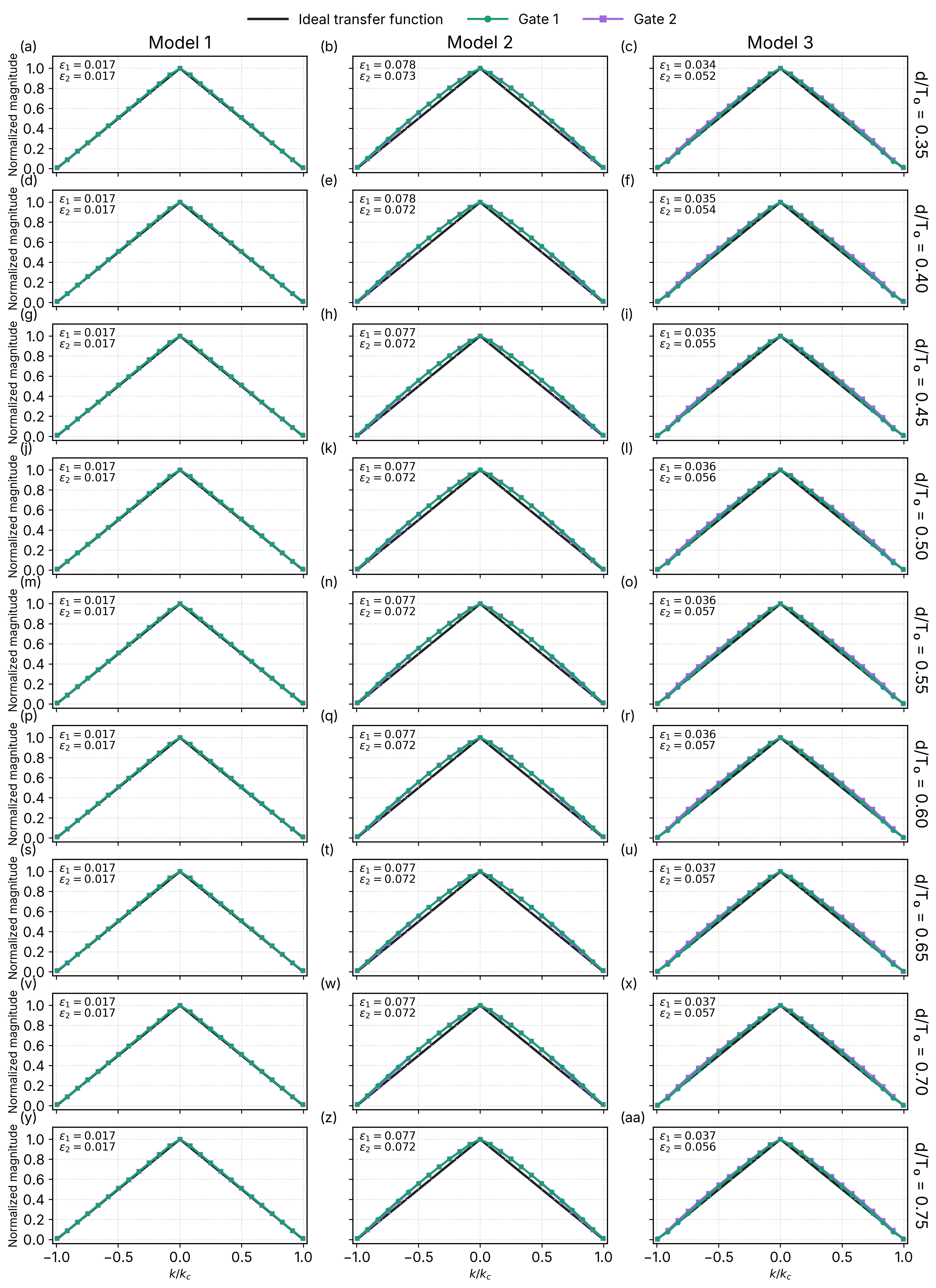}
    \caption{
    Transfer functions extracted from the 2-MOSD data used for the resolution comparison for the three FWM models (columns) and normalized peak separations from $d/T_0=0.35$ to $0.75$ (rows). The ideal transfer function (black) is compared with the extracted transfer functions of the first (green) and second (purple) gates. The corresponding relative 2-norm errors $\epsilon_1$ and $\epsilon_2$ are given in the upper-left corner of each panel.
    }
    \label{fig:fwm_27panelkernels}
\end{figure}

Figure~\ref{fig:fwm_27panelkernels} shows the corresponding extracted transfer functions. The transfer functions depend only weakly on the peak separation, indicating that the trends observed in the reconstructed distributions are not associated with substantial changes in the extracted kernels. Because nuisance is excluded from the resolution study, the additional scale ambiguity affecting the second-gate transfer function in the nuisance-inclusive reconstruction is absent here.

\clearpage

\end{document}